\documentclass[showpacs]{revtex4}
\usepackage{amsmath}
\usepackage{amssymb}
\usepackage{graphicx}
\usepackage{color}
\usepackage{dcolumn}

\begin{document}

\newcolumntype{.}{D{.}{.}{-1}}

\title{Mixed neutron-star-plus-wormhole systems:\\ Equilibrium configurations
}

\author{
Vladimir Dzhunushaliev,$^{1,2,3}$
\footnote{
Email: vdzhunus@krsu.edu.kg}
Vladimir Folomeev,$^{2,3}$
\footnote{Email: vfolomeev@mail.ru}
Burkhard Kleihaus,$^{3}$
\footnote{Email: b.kleihaus@uni-oldenburg.de }
and Jutta Kunz$^3$
\footnote{Email:  jutta.kunz@uni-oldenburg.de}
}
\affiliation{$^1$Institute for Basic Research,
Eurasian National University,
Astana, 010008, Kazakhstan
 \\ 
$^2$Institute of Physicotechnical Problems and Material Science of the NAS
of the
Kyrgyz Republic, 265 a, Chui Street, Bishkek, 720071,  Kyrgyzstan \\
$^3$Institut f\"ur Physik, Universit\"at Oldenburg, Postfach 2503
D-26111 Oldenburg, Germany
}

\begin{abstract}
We study gravitationally bound, spherically symmetric
equilibrium configurations
consisting of ordinary (neutron-star) matter
and of a phantom/ghost scalar field
which provides the nontrivial topology in the system.
For such mixed configurations,
we show the existence of static, regular, asymptotically flat
general relativistic solutions.
Based on the energy approach, we discuss the stability
as a function of the core density of the neutron matter
for various sizes of the wormhole throat.
\end{abstract}

\maketitle

\section{Introduction}

The discovery of the accelerated expansion of the present Universe
at the end of the 1990s started a new era in
our understanding of the Universe.
It became clear that besides visible and dark matter
which are gravitationally clustered in galaxies and galaxy clusters
there should exist, in addition, a fundamentally different form of energy
in the Universe -- dark energy,
accounting for about three quarters of the energy content of the Universe.
This dark energy is not gravitationally clustered,
but distributed rather uniformly in space.
Since it has a high negative pressure,
which is rather unusual from the point of view of ordinary matter,
it can provide for the current acceleration of the Universe.

The key property of dark energy lies in its ability to violate
various energy conditions.
Of greatest interest for cosmology are the following energy conditions:
(i) the strong energy condition, $(\varepsilon +3 p)\geq 0$;
(ii) the weak/null energy condition $(\varepsilon +p)\geq 0$
(where $\varepsilon$ and $p$ are the effective energy density
and the pressure of the matter fields filling the Universe).
To provide for the accelerated expansion,
it is sufficient to violate the strong energy condition.
In turn, the failure to satisfy the weak energy condition results in an
accelerated expansion of the Universe that is
even faster than exponential.

If some form of matter violating these energy conditions does indeed exist
in the present Universe, then it may serve not only
as a mechanism for providing the current acceleration.
It may as well be present in compact astrophysical objects.
Indeed, despite the fact that in describing the evolution of the Universe
it is assumed that dark energy is homogeneously distributed in the Universe,
it is possible to imagine a situation where inhomogeneities
may arise due to gravitational instabilities.
This could lead to a collapse of lumps of dark energy
with the subsequent creation of compact configurations
-- so-called dark energy stars
\cite{Mazur:2004ku,Dymnikova:2004qg,Lobo:2005uf,DeBenedictis:2005vp,
DeBenedictis:2008qm,Gorini:2008zj,Gorini:2009em,Dzhunushaliev:2008bq,
Dzhunushaliev:2011ma,Folomeev:2011aa,Yazadjiev:2011sd,
Mota:2004pa,Cai:2005ew,Debnath:2006iz,Lee:2009qq,Folomeev:2011uj}.

For dark energy stars, typically
at least the strong energy condition is violated.
At the present time, investigations of dark energy stars
are done in two main directions:
(i) the construction of stationary configurations to demonstrate
the existence of such objects and their potential stability
\cite{Mazur:2004ku,Dymnikova:2004qg,Lobo:2005uf,DeBenedictis:2005vp,
DeBenedictis:2008qm,Gorini:2008zj,Gorini:2009em,Dzhunushaliev:2008bq,
Dzhunushaliev:2011ma,Folomeev:2011aa,Yazadjiev:2011sd};
(ii) the study of
the possible formation process of such compact objects
during gravitational collapse
\cite{Mota:2004pa,Cai:2005ew,Debnath:2006iz,Lee:2009qq,Folomeev:2011uj}.

Although the true nature of dark energy is currently unknown,
several ways have been suggested to model it.
The simplest variant seems to be the assumption
that dark energy is nothing else than
Einstein's cosmological constant $\Lambda$,
which arises as the energy density of the vacuum.
However,
the well-known ``cosmological constant problem''
has led to the consideration of models
where the dark energy evolves in time.
Perhaps one of the most developed directions here
are theories with various types of scalar fields.

Scalar fields arise naturally in particle physics including string theory.
Using them, a wide variety of scalar-field dark energy models have
been proposed (for a review, see, e.g.,~\cite{sahni:2004,Copeland:2006wr}).
One of the tools developed for modeling dark energy
are the so-called phantom/ghost fields.
Their important property is that they violate
the weak/null energy conditions.
Based on such fields, in Ref.~\cite{Caldwell:1999ew}
a simple cosmological model was suggested,
where the phantom dark energy is provided by a ghost scalar field
with a negative kinetic term
(for the further development of such models,
see the review \cite{Copeland:2006wr}).

In view of such highly unusual properties of matter
for which the weak/null energy conditions are violated,
localized objects containing such matter should differ strongly
from ordinary stars as well.
A striking example of the effects of the presence of such exotic matter
is the possibility to allow for traversable wormholes --
compact configurations with a nontrivial space-time topology.
The term ``wormhole''  has appeared in the middle of the 1950s in the work of Wheeler
\cite{Wheeler:1955zz}, who suggested
the geometrical model of electric charge
in the form of a tunnel connecting two
space-time regions and filled by an electric field.
The Wheeler wormhole was not traversable. But, this idea
 has stimulated a great deal of interest in studying models with
a nontrivial space-time topology.
One of the most significant achievements in this direction is the model
 of a traversable wormhole
suggested in 1988 by Morris and Thorne \cite{Thorne:1988}
as a toy model allowing for interstellar travel.
Moreover, since recent observational data indicates
that matter violating the weak/null energy conditions
may indeed exist in the present Universe \cite{Star1},
traversable wormholes have received
increasing attention ever since as objects which could really exist in nature.

Providing a nontrivial topology, massless ghost scalar fields
were employed in the early pioneering work
of Refs.~\cite{Bronnikov:1973fh,Ellis:1973yv}
(see also earlier work by Bergmann and Leipnik \cite{Bergmann:1957},
where they found a solution for a
massless ghost scalar field,
but not dealing with the question of the wormhole interpretation of the solution obtained).
Further examples of configurations with nontrivial topology were found
in \cite{Kodama:1978dw,Kodama:1979},
based on a ghost scalar field with a Mexican hat potential.
Interestingly, it was found that this system
has only regular, stable solutions
for a topologically nontrivial (wormhole-like) geometry.
In \cite{kuhfittig,lxli,ArmendarizPicon:2002km,Sushkov:2002ef,lobo,sushkov}
traversable Lorentzian wormholes were further investigated,
refining the conditions on the type of matter fields
that would lead to such space-times.
A general overview on the subject of Lorentzian wormholes
and violations of the various energy conditions
can be found in the book of Visser \cite{Visser}.

A recent development concerning such wormhole models
is based on the expectation,
that such topologically nontrivial objects, being traversable,
could be bound to ordinary matter
satisfying all energy conditions.
In \cite{arXiv:1102.4454},
we have suggested such a model of a mixed configuration,
consisting of a wormhole
(supported by a massless ghost scalar field)
filled by a perfect polytropic fluid.
We have shown that there exist static, regular solutions
describing mixed star-plus-wormhole configurations
which possess new physical properties
that distinguish them from ordinary stars.
A preliminary stability analysis
performed only for the external region of the configuration suggested
that those solutions could be stable with respect to linear perturbations.

The objective of the present paper is to continue the study
of the influence of a wormhole
on the structure and the physical properties of compact stars
made otherwise from ordinary matter.
To provide a nontrivial topology in this model,
we here employ a massless ghost scalar field
as one of the simplest possibilities allowing for a nontrivial topology.
For the ordinary matter filling the wormhole,
we choose neutron matter modeled within the polytropic approximation
[in the form of Eq.~\eqref{eqs_NS_WH}].

The paper is organized as follows:
In Sec.~\ref{gen_equations_WH_NS},
the general set of equations is derived,
describing equilibrium configurations
consisting of a massless ghost scalar field
and ordinary matter approximated by a polytropic equation of state.
Here, the boundary conditions are also discussed.
In Sec.~\ref{static_solutions_WH_NS},
we discuss the physical properties of the solutions,
the radii, the masses, the binding energies and the pressures.
We then present the numerically obtained static solutions
for two sets of polytropic parameters
and consider the issue of their stability within the energy approach.
Finally, in Sec.~\ref{conclus_WH_NS}
we summarize the results obtained
and address possible observable effects
which may follow from the model considered.

\section{Derivation of the equations for equilibrium configurations}
\label{gen_equations_WH_NS}

\subsection{Lagrangian and general set of equations}

We consider a model of a gravitating massless ghost scalar field
in  the  presence of a perfect fluid.
Our starting point is the Lagrangian
\begin{equation}
\label{lagran_wh_star_poten}
L=-\frac{c^4}{16\pi G}R-\frac{1}{2}\partial_{\mu}\varphi\partial^{\mu}\varphi + L_m~.
\end{equation}
Here, $\varphi$ is the ghost scalar field,
$L_m$ is the Lagrangian of the perfect isotropic fluid
(where isotropic means that the radial and the tangential pressure
of the fluid agree)
which has the form $L_m=p$ \cite{Stanuk1964,Stanuk}.
This Lagrangian leads to the corresponding energy-momentum tensor
 \begin{equation}
\label{emt_wh_star_poten}
T_i^k=(\varepsilon+p)u_i u^k-\delta_i^k p-\partial_{i}\varphi\partial^{k}\varphi
+\frac{1}{2}\delta_i^k\partial_{\mu}\varphi\partial^{\mu}\varphi,
\end{equation}
where $\varepsilon$ and $p$ are the energy density
and the pressure of the fluid, $u^i$ is the four-velocity.
For our purposes, it is convenient to choose the static metric in the form
\begin{equation}
\label{metric_wh_poten}
ds^2=e^{\nu} c^2 dt^2-e^{\lambda}dr^2-r^2d\Omega^2,
\end{equation}
where $\nu$ and $\lambda$ are functions of the radial coordinate $r$ only,
and $d\Omega^2$ is the metric on the unit 2-sphere.
Also, let us present the metric in Schwarzschild-type coordinates
which are frequently used in modeling wormholes
\begin{equation}
\label{metric_wh_sch}
ds^2=e^{\nu(r)} c^2 dt^2-\frac{dr^2}{1-b(r)/r}-r^2 d\Omega^2.
\end{equation}
This parametrization of the metric is employed
later on
to derive the boundary conditions at the
wormhole throat at the core of the configuration.

The $(_0^0)$ and $(_1^1)$ components of the Einstein equations
for the metric \eqref{metric_wh_poten}
and the energy-momentum tensor \eqref{emt_wh_star_poten}
are then given by
\begin{eqnarray}
\label{Einstein-00_poten}
&&G_0^0=-e^{-\lambda}\left(\frac{1}{r^2}
-\frac{\lambda^\prime}{r}\right)+\frac{1}{r^2}
=\frac{8\pi G}{c^4} T_0^0,
 \\
\label{Einstein-11_poten}
&&G_1^1=-e^{-\lambda}\left(\frac{1}{r^2}
+\frac{\nu^\prime}{r}\right)+\frac{1}{r^2}
=\frac{8\pi G}{c^4} T_1^1,
\end{eqnarray}
where a ``prime'' denotes differentiation with respect to $r$.

The field equation for the scalar field
is obtained by varying the Lagrangian
\eqref{lagran_wh_star_poten} with respect to $\varphi$,
\begin{equation}
\label{sf_eq_gen}
\frac{1}{\sqrt{-g}}\frac{\partial}{\partial x^i}
\left[\sqrt{-g}g^{ik}\frac{\partial \varphi}{\partial x^k}\right]= 0.
\end{equation}
Using the metric
 \eqref{metric_wh_poten}, this equation is integrated to give
\begin{equation}
\label{sf_poten}
\varphi^{\prime 2}=\frac{D^2}{r^4}e^{\lambda-\nu},
\end{equation}
where $D$ is an integration constant.

Not all of the Einstein field equations are independent
because of the law of conservation of energy and momentum
$T^k_{i;k}=0$.
Taking the $i=1$ component of this equation gives
\begin{equation}
\label{conserv_1}
\frac{d T^1_1}{d r}+
\frac{1}{2}\left(T_1^1-T_0^0\right)\nu^\prime+\frac{2}{r}\left[T_1^1-\frac{1}{2}\left(T^2_2+T^3_3\right)\right]=0.
\end{equation}
The remaining two Einstein equations are then satisfied as a
consequence of the Eqs.~(\ref{Einstein-00_poten}),
(\ref{Einstein-11_poten}), and (\ref{conserv_1}).

Thus, we have five unknown functions:
$\nu, \lambda, \varphi, \varepsilon$, and $p$.
Keeping in mind that $\varepsilon$ and $p$ are related by an equation of state,
there are only four unknown functions.
For these functions, there are four equations:
the two Einstein equations \eqref{Einstein-00_poten} and \eqref{Einstein-11_poten},
the scalar-field equation \eqref{sf_poten},
and the equation of hydrostatic equilibrium \eqref{conserv_1}.
Using the energy-momentum tensor \eqref{emt_wh_star_poten},
the right-hand sides of the Eqs.~\eqref{Einstein-00_poten} and \eqref{Einstein-11_poten}
take the form
\begin{eqnarray}
\label{T00_stat}
&&T_0^0=\varepsilon-\frac{1}{2}e^{-\lambda}\varphi^{\prime 2},
 \\
\label{T11_stat}
&&T_1^1=- p+\frac{1}{2}e^{-\lambda}\varphi^{\prime 2}.
\end{eqnarray}

Then, taking into account the expressions
for components of the energy-momentum tensor \eqref{emt_wh_star_poten}
$$
T_2^2=T_3^3=- p-\frac{1}{2}e^{-\lambda}\varphi^{\prime 2},
$$
and using \eqref{sf_poten},
we obtain from \eqref{conserv_1} the following equation for
hydrostatic equilibrium,
\begin{equation}
\label{conserv_2}
\frac{d p}{d r}=-\frac{1}{2}(\varepsilon+p)\frac{d\nu}{d r}.
\end{equation}

\subsection{Equation of state}

To model the neutron matter filling the wormhole,
it is necessary to choose an appropriate equation of state.
The choice of the equation of state plays a crucial role
in modeling the neutron-star structure.
A large number of neutron matter equations of state
have been suggested, providing different sets of densities and pressures
for the neutron stars.
These include a parametric dependence as
in the pioneering work of Oppenheimer and Volkoff \cite{Oppen1939},
a polynomial dependence as in \cite{Cameron1959},
an equation of state obtained
from field theoretical considerations as in \cite{DAl1985},
or a unified equation of state
(see, e.g., Ref.~\cite{Haensel:2004nu}
where the analytical representations of unified equations of state of
neutron-star matter are given).

Here, we employ a simplified variant for the equation of state,
where a more or less realistic neutron matter equation of state
is approximated in the form of a polytropic equation of state.
In particular, we employ the following parametric relation
between the pressure and the energy density of the fluid,
$$
\varepsilon=n_b m_b c^2
+ \frac{p}{\gamma-1} , \quad
p=k c^2 n_{b}^{(ch)} m_b \left(\frac{n_b}{n_{b}^{(ch)}}\right)^\gamma,
$$
where $n_{b}$ is the baryon number density,
$n_{b}^{(ch)}$ is some characteristic value of $n_{b}$,
$m_b$ is the baryon mass,
and $k$ and $\gamma$ are parameters
whose values depend on the properties of the neutron matter.

For our purpose it is convenient to rewrite the above equation of state
in the form
\begin{equation}
\label{eqs_NS_WH}
p=K \rho_{b}^{1+1/n}, \quad \varepsilon = \rho_b c^2 +n p,
\end{equation}
with the constant $K=k c^2 (n_{b}^{(ch)} m_b)^{1-\gamma}$,
the polytropic index $n=1/(\gamma-1)$,
and $\rho_b=n_{b} m_b$ denotes the rest-mass density
of the neutron fluid.

Setting
$m_b=1.66 \times 10^{-24}\, \text{g}$
and $n_{b}^{(ch)} = 0.1\, \text{fm}^{-3}$,
we consider below configurations with two sets
of values for the parameters $k$ and $\gamma$:
\begin{itemize}
\item[(i)]
$k=0.0195$ and $\gamma=2.34$ \cite{Damour:1993hw},
adjusted to fit the equation of state II
of Ref.~\cite{DAl1985};
we denote this choice by EOS1 in this paper.
\item[(ii)]
$k=0.1$ and $\gamma=2$  \cite{Salg1994},
corresponding to a gas of baryons interacting
via a vector meson field, as described by Zel'dovich \cite{Zeld1961,Zeld}
(see also Ref.~\cite{Tooper2}
where relativistic configurations
with such an equation of state were considered);
we denote this choice by EOS2 in this paper.
\end{itemize}

Introducing the new variable $\theta$ \cite{Zeld}
\begin{equation}
\label{theta_def}
\rho_b=\rho_{b c} \theta^n~,
\end{equation}
where $\rho_{b c}$ is the density of the neutron fluid at the
wormhole throat
(or the center of the star in the case without a wormhole)
we may rewrite the pressure and the energy density,
Eq.~\eqref{eqs_NS_WH}, in the form
\begin{equation}
\label{pressure_fluid_theta}
p=K\rho_{b c}^{1+1/n} \theta^{n+1}, \quad
\varepsilon =  \left( \rho_{b c} c^2 +
  n K \rho_{b c}^{1 + {1}/{n} } \theta \right) \theta^n.
\end{equation}
Making use of this expression,
we obtain
for the internal region with $\theta \ne 0$
from Eq.~\eqref{conserv_2}
\begin{equation}
\label{conserv_3}
2\sigma(n+1)\frac{d\theta}{d r}=
-\left[1+\sigma(n+1) \theta\right]\frac{d\nu}{dr},
\end{equation}
where $\sigma=K \rho_{b c}^{1/n}/c^2=p_c/(\rho_{b c} c^2)$ is a constant,
related to the pressure $p_c$ of the fluid at the wormhole throat
(or at the center of the star in the case without a wormhole).
This equation may be integrated to
give in the internal region with $\theta \ne 0$ the metric function $e^{\nu}$
in terms of $\theta$,
\begin{equation}
\label{nu_app}
e^{\nu}=e^{\nu_c}\left[\frac{1+\sigma (n+1)}{1+\sigma (n+1)\theta}\right]^{2},
\end{equation}
and $e^{\nu_c}$ is the value of $e^{\nu}$ at the throat where $\theta=1$.
The integration constant $\nu_c$ is fixed
by requiring that the space-time is asymptotically flat,
i.e., $e^{\nu}=1$ at infinity.

\subsection{Internal set of equations}

We first consider the set of equations in the
internal region, where $\theta \neq 0$.
Here, the system is characterized by three unknown functions:
$\lambda$, $\theta$ and $\varphi$.
These three functions are determined by the three Eqs.
\eqref{Einstein-00_poten}, \eqref{Einstein-11_poten}, and \eqref{sf_poten},
and also by the relation \eqref{nu_app}.
It is convenient to rewrite these equations
by introducing the new function $M(r)$,
\begin{equation}
\label{u_app}
e^{-\lambda}=1-\frac{2 G M(r) }{c^2 r}\,.
\end{equation}
The function $M(r)$ can be interpreted as the total mass within
the areal radius $r$.
Thus, for $r=r_b$,
where $r_b$ denotes the outer boundary of the fluid where $\theta=0$
(for more details on this, see Sec.~\ref{static_solutions_WH_NS}),
we obtain the total mass of the configuration within
the boundary of the star of radius $r_b$.
The limit $r \to \infty$ then yields the total mass of the configuration.

With this function $M(r)$, Eq.~\eqref{Einstein-00_poten} yields
\begin{equation}
\label{00_via_u}
\frac{d M}{d r}=\frac{4\pi}{c^2}\, r^2\left[\varepsilon-\frac{1}{2}\left(1-\frac{2 G M }{c^2 r}\right)\varphi^{\prime 2}\right].
\end{equation}
For the spherically symmetric case without a wormhole
(corresponding in our case to the absence of the scalar field),
we need to require the boundary condition $M(0)=0$
in order to guarantee regularity at the origin \cite{Tooper:1964}.
This corresponds to the fact that there is no mass
associated with the origin, $r=0$.
For a wormhole, on the other hand,
there exists a finite minimal value of the radius, the throat $r=r_0$,
and this is associated with a finite value of the mass function at the
throat, $M(r_0)\neq 0$.

Now, we introduce dimensionless variables
\begin{equation}
\label{dimless_xi_v}
\xi=A r, \quad v(\xi)=\frac{A^3 M(r)}{4\pi \rho_{b c}},
\quad \phi=\left[\frac{4\pi G}{\sigma(n+1)c^4}\right]^{1/2}\varphi,
\quad \text{with} \quad
A=\left[\frac{4\pi G \rho_{b c}}{\sigma(n+1)c^2}\right]^{1/2},
\end{equation}
where $A$ has the dimension of inverse length, and rewrite
Eqs.~\eqref{00_via_u} and \eqref{Einstein-11_poten} in the form
\begin{eqnarray}
\label{eq_v_app}
&&\frac{d v}{d\xi}=\xi^2\left\{(1+n\sigma \theta)\theta^n
-\frac{1}{2}\frac{\bar{D}^2}{\xi^4}e^{-\nu_c}\left[\frac{1+\sigma (n+1)\theta}{1+\sigma (n+1)}\right]^{2}
\right\},
\\
\label{eq_theta_app}
&&\xi^2\frac{1-2\sigma(n+1)v/\xi}{1+\sigma(n+1)\theta}\frac{d\theta}{d\xi}=
\xi^3\left\{\theta^n\left[1+\sigma(n-1)\theta\right]-\frac{1}{\xi^2}\frac{d v}{d\xi}\right\}-v,
\end{eqnarray}
where we have used expression \eqref{sf_poten}
and introduced the dimensionless constant
$$
\bar{D}=\frac{4\pi G D}{\sigma(n+1) c^3} \sqrt{\rho_{b c}}~.
$$

Thus, the internal part ($r\le r_b$)
of the static configurations under consideration
is described by the Eqs.~\eqref{eq_v_app} and \eqref{eq_theta_app}
together with the scalar-field equation
\begin{equation}
\label{eq_phi_app}
        \left(\frac{d\phi}{d\xi}\right)^2 =
        \frac{\bar{D}^2}{\xi^4}
        \frac{e^{-\nu_c}}{1-2\sigma(n+1)v/\xi}
        \left[\frac{1+\sigma (n+1)\theta}{1+\sigma (n+1)}\right]^{2}.
\end{equation}

\subsection{External set of equations}

Let us next turn to the external part ($r\ge r_b$) of the solutions
outside the fluid.
Asymptotic flatness of the external solutions requires
that the metric function $e^{\nu}$ tends to one at infinity.
This in turn
determines the value of the integration constant $\nu_c$ at the throat.

To find the external solutions, we start again from the Einstein equations
\eqref{Einstein-00_poten} and \eqref{Einstein-11_poten},
and the scalar-field equation \eqref{sf_poten},
taking into account that in the external region there is no ordinary matter,
i.e., $\varepsilon=p=0$.
This leads to the following system of equations,
\begin{eqnarray}
\label{Einstein-00_ext}
&&-e^{-\lambda}\left(\frac{1}{r^2}-\frac{\lambda^\prime}{r}\right)
+\frac{1}{r^2}
=-\frac{4\pi G}{c^4}\frac{D^2}{r^4} e^{-\nu},
\\
\label{Einstein-11_ext}
&&-e^{-\lambda}\left(\frac{1}{r^2}+\frac{\nu^\prime}{r}\right)+\frac{1}{r^2}
=\frac{4\pi G}{c^4}\frac{D^2}{r^4} e^{-\nu},\\
\label{first_int_ext}
&&\varphi^{\prime 2}=\frac{D^2}{r^4}e^{\lambda-\nu}.
\end{eqnarray}
These can be rewritten in terms of the dimensionless variables
$v(\xi)$, $\phi(\xi)$ and $\nu(\xi)$ as follows,
\begin{eqnarray}
\label{eq_v_ext}
	\frac{d v}{d\xi} &=& -\frac{1}{2}\frac{\bar{D}^2}{\xi^2}e^{-\nu},
\\
\label{eq_nu_ext}
	\frac{d\nu}{d\xi} &=&
	\frac{1}{\xi}\left[
	\frac{1-\sigma (n+1)\frac{\bar{D}^2}{\xi^2}e^{-\nu}}
                {1-2\sigma (n+1) v/\xi} -1\right],
\\
\label{eq_phi_ext}
	\left(\frac{d\phi}{d\xi}\right)^2 &=&
	\frac{\bar{D}^2}{\xi^4}
	\frac{e^{-\nu}}{1-2\sigma (n+1) v/\xi}\,.
\end{eqnarray}
This system still contains the parameter $\sigma$ as a trace
of the influence of the fluid on the external solution
due to the definition of the dimensionless quantities (\ref{dimless_xi_v}).

\subsection{Expansion at the throat}

Finally, we need to consider the appropriate boundary conditions
at the wormhole throat corresponding to the core of the configurations
\cite{arXiv:1102.4454}.

Let us here briefly comment on the nomenclature used.
In contrast to the case of ordinary stars
having a center at the point $r = 0$,
there exists a finite minimal value of the radial coordinate $r = r_{0}$
corresponding to the radius of the throat
in the case of the star-plus-wormhole systems considered here.
Moreover, because of the presence of ordinary matter,
the configurations differ also from simple wormholes.
To take this into account, we employ the term ``core''
to describe the ``throat'' area of such star-plus-wormhole systems.

In terms of the dimensionless variables introduced above,
$\xi=\xi_0$ denotes the coordinate at the core of the configuration.
Here, the dimensionless density of the fluid is characterized by
\begin{equation}
\label{bound_theta}
\theta_0\equiv\theta(\xi_0)=1.
\end{equation}
This condition corresponds to the fact
that at the wormhole throat the density of the fluid is $\rho_{b c}$.

In order to determine the initial value of the function $v$,
we consider that the metric \eqref{metric_wh_sch}
satisfies the condition $(1-b(r)/r)\geq 0$
throughout the space-time \cite{Thorne:1988,Visser}.
This implies
\begin{equation}
\label{cond_wh}
b(r_0)=r_0, \quad b^\prime(r_0)<1,
\quad \text{and} \quad b(r)<r
\quad \text{for} \quad r > r_0.
\end{equation}
Comparing the form of the metric \eqref{metric_wh_sch}
with the form of the metric \eqref{metric_wh_poten}
used for the derivation of the
Eqs.~\eqref{eq_v_app} and \eqref{eq_theta_app},
and taking into account expression \eqref{u_app},
together with the dimensionless variables \eqref{dimless_xi_v},
we find
\begin{equation}
\label{cond_b}
\frac{b(r)}{r}=\frac{2 G M(r) }{c^2 r}
\quad \Rightarrow  \quad \frac{B(\xi)}{\xi}=2\sigma(n+1)\frac{v}{\xi},
\end{equation}
where we have introduced the dimensionless function $B(\xi)\equiv A b$.
From this expression, taking into account the relation
$B(\xi_0)/\xi_0=1$ from \eqref{cond_wh},
we obtain the boundary condition for the function $v$ at the throat
\begin{equation}
\label{bound_v}
v_0\equiv v(\xi_0)=\frac{1}{2\sigma(n+1)}\,\xi_0.
\end{equation}

This boundary condition implies that, as $\xi \to \xi_0$,
the following expression vanishes,
\begin{equation}
\label{zero_expr}
\left[1-2\sigma(n+1)\frac{v}{\xi}\right] \to 0.
\end{equation}
This in turn means that the coefficient
in front of the derivative of the function $\theta$
in Eq.~\eqref{eq_theta_app} goes to zero,
which leads to a singularity in a generic solution.

To obtain solutions, which are regular at the throat
we consider a Taylor series expansion of the functions $v$ and $\theta$
in the neighborhood of the point $\xi=\xi_0$.
At a point $\xi=\xi_1$ close to the throat the expansion for $v$ reads
to first order
\begin{equation}
\label{v_series}
v(\xi_1)=v_0+v_1(\xi_1-\xi_0).
\end{equation}
Substituting this into Eq.~\eqref{eq_v_app},
we obtain for $v_1$ the following expression,
\begin{equation}
\label{const_v1}
v_1= \xi_0^2\left\{(1+n\sigma \theta_0)\theta_0^n
-\frac{1}{2}\frac{\bar{D}^2}{\xi_0^4}e^{-\nu_c}
\left[\frac{1+\sigma (n+1)\theta_0}{1+\sigma (n+1)}\right]^{2}\right\}.
\end{equation}

Regularity of the solutions of Eq.~\eqref{eq_theta_app} is achieved,
when we assume that, together with \eqref{zero_expr},
the right-hand side of \eqref{eq_theta_app}
goes simultaneously to zero.
Proceeding from this requirement,
we obtain the following expression for $\bar{D}$:
\begin{equation}
\label{const_D}
\bar{D}^2=2\xi_0^4 e^{\nu_c}
\left[\frac{1+\sigma (n+1)}{1+\sigma (n+1)\theta_0}\right]^{2}
\left(\sigma \theta_0^{n+1}+\frac{v_0}{\xi_0^3}\right).
\end{equation}
Substituting this into \eqref{const_v1}, we finally obtain
\begin{equation}
\label{const_v1_f}
v_1= \xi_0^2\left\{ \theta_0^n\left[1+\sigma(n-1)\theta_0\right]
     -\frac{v_0}{\xi_0^3}\right\}.
\end{equation}
The Taylor series expansion for the function $\theta$ at
at the point $\xi=\xi_1$
reads to first order
\begin{equation}
\theta(\xi_1) = 1 +\theta_1
            (\xi_1-\xi_0)~, 
\label{theta_exp}
\end{equation}
where we obtain for the coefficient $\theta_1$,
\begin{equation}
\theta_1 =\xi_0\frac{[1+\sigma (n+3)][1+\sigma (n+1)]}
                 {\xi_0^2 n\sigma (1+n\sigma)-[1-\xi_0^2\sigma (1-\sigma)]}~.
\label{theta_1}
\end{equation}
Nonsymmetric wormholes
(with respect to the two asymptotically flat space-times)
could also be obtained,
but these would satisfy a different set of boundary conditions.

Thus, a static equilibrium solution is obtained as follows:
We start the numerical integration at the point $\xi=\xi_1$,
solving numerically
the system of equations \eqref{eq_v_app} and \eqref{eq_theta_app}
subject to the boundary conditions \eqref{v_series} with \eqref{bound_v} and
\eqref{const_v1_f} for $v$,
and 
\eqref{theta_exp} with \eqref{theta_1} for $\theta$.
We then proceed with the integration until we reach the point $\xi=\xi_b$,
where the function $\theta$ becomes zero \cite{arXiv:1102.4454}.
Here, the energy density associated with the fluid vanishes.
The surface bounded by the radius $\xi=\xi_b$ represents
the boundary of the neutron matter.
Note, that the energy density associated with the scalar field
is, in general, still finite at this boundary.

Now, the external part of the solution is sought,
starting from the surface of the fluid at $\xi=\xi_b$
with the boundary conditions $v(\xi_b)$ and $\nu(\xi_b)$
as determined from the internal part of the solution.
Requiring asymptotical flatness of the space-time
finally allows to determine the value of the integration constant $\nu_c$
from \eqref{nu_app}
by requiring $e^{\nu}$ to be equal to unity at infinity.
(The values of $\nu_c$ for the examples shown in
Fig.~\ref{energ_metr_fig} are given in the caption.)
Thus, the complete solution for the configuration under consideration is derived
by matching of the internal fluid solutions given by Eqs.~\eqref{eq_v_app}-\eqref{eq_phi_app}
with the external solutions obtained from the system \eqref{eq_v_ext}-\eqref{eq_phi_ext}.

\section{Static solutions
}
\label{static_solutions_WH_NS}

In this section, we discuss the numerical solutions of the above
sets of internal and external equations and their physical properties.

\subsection{Radii}

The radial coordinate $r$ describes the areal radius
of a sphere with area $4 \pi r^2$.
The throat radius corresponds to the areal radius $r_0$,
and the neutron matter of the star is contained
within the areal radius $r_b$ (denoted $R$ in the tables).
The gravitational radius of the system corresponds to the
areal radius $r_g= 2 G M/c^2$, where $M$ is the total mass.
In dimensionless coordinates, the areal radius is given
in terms of the coordinate $\xi$.
Thus, $\xi_0$ denotes the throat radius and
$\xi_b$ the radius of the neutron fluid.

Another physically relevant radial coordinate is given
by the coordinate $\bar{\xi}$
associated with the proper radius,
which gives the distance from the throat.
It is defined as follows,
$$
\bar{\xi}=\int_{\xi_0}^{\xi} e^{\lambda/2}d\xi,
$$
or, taking into account
Eqs.~\eqref{u_app} and \eqref{dimless_xi_v},
\begin{equation}
\label{xi_observ}
\bar{\xi}=\int_{\xi_0}^{\xi} \left[1-2\sigma(n+1)
\frac{v(\xi')}{\xi'}\right]^{-1/2}d\xi'.
\end{equation}
Then, the proper radius of the fluid $R_{\text{prop}}$ is obtained
in dimensional variables as $R_{\text{prop}}=\bar{\xi}_{b}/A$.

\subsection{Mass contributions}
\label{energ_mass}

The system under consideration consists of two parts:
the internal region $r_0 \le r \le r_b$
and the external region $r_b \le r \le \infty$.
Correspondingly, the energy density of the system
is given by the internal energy density, $\varepsilon_{\text{int}}$
obtained from the expressions
\eqref{T00_stat}, \eqref{eqs_NS_WH},
\eqref{theta_def}, \eqref{nu_app}, and \eqref{dimless_xi_v}
\begin{equation}
\label{dens_int}
\varepsilon_{\text{int}} \equiv
\left[T_0^0\right]_{\text{int}}=\rho_{b c} c^2
\left\{(1+n\sigma \theta)\theta^n
-\frac{1}{2}\frac{\bar{D}^2}{\xi^4}e^{-\nu_c}
\left[\frac{1+\sigma (n+1)\theta}{1+\sigma (n+1)}\right]^{2} \right\}
\end{equation}
and the external energy density, $\varepsilon_{\text{ext}}$,
\begin{equation}
\label{dens_ext}
\varepsilon_{\text{ext}} \equiv
\left[T_0^0\right]_{\text{ext}}=-\frac{\rho_{b c} c^2}{2}
\frac{\bar{D}^2}{\xi^4}e^{-\nu} .
\end{equation}
The total energy density of the configuration is
\begin{equation}
\varepsilon_t=
\varepsilon_{\text{int}} \Theta(\xi_b-\xi)
+\varepsilon_{\text{ext}} \Theta(\xi-\xi_b).
\label{total_edens}
\end{equation}

As discussed above, the effective mass $M(r)$ inside
a surface with radius $r$
is given by \cite{Thorne:1988,Visser}
$$
M(r) = \frac{c^2}{2 G}\,b(r)
=\frac{c^2}{2 G}\,r_0+\frac{4\pi}{c^2}\int_{r_0}^{r}
\varepsilon_t (r') r'^{2} dr',
$$
where the integration constant is determined by $b(r_0)=r_0$,
Eq.~\eqref{cond_wh}.
Employing dimensionless variables \eqref{dimless_xi_v}
and $B(\xi)$ [see Eq.~\eqref{cond_b}]
we obtain the dimensionless effective mass ${\cal M}(\xi)$,
\begin{eqnarray}
\label{dimls_mass}
{\cal M}(\xi)\equiv \frac{B(\xi)}{2}=
\xi_0/2 &+&\sigma(n+1) \int_{\xi_0}^{\xi}
\left\{ (1+n\sigma \theta)\theta^n
-\frac{1}{2}\frac{\bar{D}^2}{\xi'^4}e^{-\nu_c}
\left[\frac{1+\sigma (n+1)\theta}{1+\sigma (n+1)}\right]^{2}
\right\}  \Theta(\xi_b-\xi') \xi'^2 d\xi'
\nonumber \\
&-& \frac{\sigma(n+1)\bar{D}^2}{2} \int_{\xi_0}^{\xi}
 \frac{e^{-\nu}}{\xi'^2} \Theta(\xi'-\xi_b) d\xi'
\end{eqnarray}
with
$ M(r) = M^* {\cal M}(\xi) $, where the quantity
$$M^*=\sqrt{\frac{K(n+1)}{4\pi G^3}}\rho_{b c}^{\gamma/2 -1} c^2 $$
fixes the scale of the mass.

The asymptotic value $\lim\limits_{r \to \infty}M(r) = M$
corresponds to the total mass of the configuration,
while in dimensionless units
$\lim\limits_{\xi \to \infty}{\cal M}(\xi) = {\cal M}$.
For later reference, we now subdivide this expression
for the total mass into four dimensionless components according to
\begin{equation}
\label{dim_mass}
{\cal M}=
{\cal M}_{\text{th}}+{\cal M}_{\text{fl}}+
{\cal M}_{\text{sfint}}+{\cal M}_{\text{sfext}}
\end{equation}
with the mass at the throat
$${\cal M}_{\text{th}}=\xi_0/2;$$
the mass of the fluid
$${\cal M}_{\text{fl}}=\sigma(n+1) \int_{\xi_0}^{\xi_b}
(1+n\sigma \theta)\theta^n \xi'^2 d\xi';$$
the internal part of the mass of the scalar field
$${\cal M}_{\text{sfint}}=-\frac{\sigma(n+1)\bar{D}^2
e^{-\nu_c}}{2} \int_{\xi_0}^{\xi_b}
\frac{1}{\xi'^2}\left[\frac{1+\sigma (n+1)\theta}{1+\sigma (n+1)}\right]^{2}
d\xi';$$
and the external part of the mass of the scalar field
$${\cal M}_{\text{sfext}}=-\frac{\sigma(n+1)\bar{D}^2}{2} \int_{\xi_b}^{\infty}
\frac{e^{-\nu}}{\xi'^2}\, d\xi'.$$

\subsection{Binding energy}

We start from the total energy $E$ of the system given by
\begin{equation}
\label{total_energ}
E=M c^2= M^* \Big(
{\cal M}_{\text{th}}+{\cal M}_{\text{fl}}
+{\cal M}_{\text{sfint}}+{\cal M}_{\text{sfext}}
\Big) c^2
= \Big( M_{\text{th}}+M_{\text{fl}}+M_{\text{sfint}}+M_{\text{sfext}}
\Big) c^2~.
\end{equation}
In order to derive a physically motivated expression for the binding energy
of the system, let us consider the relativistic continuity equation
\begin{equation}
 (n_b u^\mu)_{;\mu} = 0,
\end{equation}
with the baryon number density of the neutron fluid $n_b$
and the four-velocity $u^\mu$.
It follows that the neutron particle number $N$ is given by
\begin{equation}
N=\int_{r_0}^{r_b}{(n_b u^0) \sqrt{-g} d^3x}
   = \frac{ 4 \pi}{m_b} \int_{r_0}^{r_b}{ {\rho_b} e^{\lambda/2}r^2 dr} ,
\end{equation}
with $u^0 = e^{-\nu/2}$,
and the factor $\sqrt{-g}$ enters because the natural volume element
on the spacelike hypersurfaces is needed \cite{Harrison:1965,Wald:1984rg}.
The associated energy of $N$ free neutrons is given by
\begin{equation}
E_{fb} = N m_b c^2.
\label{Efb}
\end{equation}

For a simple neutron star without a wormhole,
the binding energy (B.E.) is defined as the difference
of the energy of $N$ free particles $E_{fb}$, Eq.~(\ref{Efb}),
and the total energy $E$, Eq.~(\ref{total_energ})
\begin{equation}
\label{bind_enrg}
\text{B.E.}=E_{fb}-E,
\end{equation}
as discussed, e.g.,~in Ref.~\cite{Zeld}. 
An analogous definition holds for a boson star
composed of $N$ massive bosons.
Indeed, in order to disperse the particles to infinity,
one has to supply precisely this amount of energy to the system.

For the combined star-plus-wormhole system,
on the other hand, we still have to consider how to deal
with the massless ghost scalar field
that supports the wormhole.
Thus, let us first address an isolated wormhole
made from a massless ghost scalar field
without any neutron matter.
With the above set of boundary conditions,
such a wormhole has zero energy (respectively, zero mass) \cite{Visser}.
This agrees with the energy of flat space in the absence
of a wormhole.

Let us therefore consider the binding energy of the star-plus-wormhole
system as the difference between the following contributions:
The energy of $N$ free particles
dispersed to infinity with a (therefore) vanishing ghost scalar field
and the energy of the combined star-plus-wormhole system.
Alternatively, we could consider
the difference between the energy of $N$ free particles
dispersed to infinity together with the energy of an isolated wormhole
and the energy of the combined star-plus-wormhole system.
Both cases yield the same result, namely
the above expression \eqref{bind_enrg},
obtained without a wormhole.

\begin{table}
 \caption{Characteristics of a set of configurations for EOS1
for several values of the dimensionless throat radius $\xi_0$.
The radius of the throat $R_{\text{th}}$,
the areal radius of the fluid $R$,
the proper radius of the fluid $R_{\text{prop}}$
and the gravitational radius $r_g$ (last column)
are given in kilometers.
The total mass $M$, the mass at the throat $M_{\text{th}}$,
the mass of the fluid $M_{\text{fl}}$,
the internal part of the mass of the scalar field $M_{\text{sfint}}$,
and the external part of the mass of the scalar field $M_{\text{sfext}}$
are given in solar mass units.
}
\vspace{.3cm}
\begin{tabular}{p{1.5cm}p{1.5cm}p{1.5cm}p{1.5cm}p{1.5cm}p{1.5cm}p{1.5cm}p{1.5cm}p{1.5cm}p{1.5cm}}
\hline \\[-5pt]
$\rho_{b c}, \text{g cm}^{-3}$ &  $ \hphantom{xx}R_{\text{th}},$ km & $ \hphantom{xx}R,$ km &$ R_{\text{prop}},$ km &
 $\hphantom{xx} M/M_\odot$ &  $\hphantom{xx} M_{\text{th}}/M_\odot$&  $\hphantom{xx} M_{\text{fl}}/M_\odot$
&  $M_{\text{sfint}}/M_\odot$&  $M_{\text{sfext}}/M_\odot$& $ \hphantom{xx} r_g,$ km \\[2pt]
\hline \\[-7pt]
\multicolumn{10}{c}{Without a wormhole} \\[2pt]
\end{tabular}
\begin{tabular}{p{1.5cm}.........}
\hline \\[-15pt]
1.0$\times 10^{14}$&	-&	12.3047&	12.4312&	0.1604&	-&	0.1604&	-&	-&	0.4732\\
2.0$\times 10^{14}$&	-&	13.3411&	13.6760&	0.4093&	-&	0.4093&	-&	-&	1.2073\\
3.0$\times 10^{14}$&	-&	13.7197&	14.2885&	0.6680&	-&	0.6680&	-&	-&	1.9704\\
4.0$\times 10^{14}$&	-&	13.8078&	14.6132&	0.9080&	-&	0.9080&	-&	-&	2.6781\\
6.0$\times 10^{14}$&	-&	13.5903&	14.8356&	1.2965&	-&	1.2965&	-&	-&	3.8241\\
8.0$\times 10^{14}$&	-&	13.1647&	14.7824&	1.5647&	-&	1.5647&	-&	-&	4.6152\\
1.0$\times 10^{15}$&	-&	12.6865&	14.6083&	1.7387&	-&	1.7387&	-&	-&	5.1286\\
1.5$\times 10^{15}$&	-&	11.5667&	14.0149&	1.9264&	-&	1.9264&	-&	-&	5.6822\\
1.8$\times 10^{15}$&	-&	11.0034&	13.6572&	1.9495&	-&	1.9495&	-&	-&	5.7504\\
2.0$\times 10^{15}$&	-&	10.6733&	13.4338&	1.9466&	-&	1.9466&	-&	-&	5.7418\\
3.0$\times 10^{15}$&	-&	9.4529&	12.5362&	1.8482&	-&	1.8482&	-&	-&	5.4516\\
6.0$\times 10^{15}$&	-&	7.9586&	11.3502&	1.5434&	-&	1.5434&	-&	-&	4.5523\\
1.0$\times 10^{16}$&	-&	7.5888&	11.1383&	1.3675&	-&	1.3675&	-&	-&	4.0336\\
1.5$\times 10^{16}$&	-&	7.7045&	11.3870&	1.3082&	-&	1.3082&	-&	-&	3.8588\\
3.0$\times 10^{16}$&	-&	8.1252&	12.0257&	1.3375&	-&	1.3375&	-&	-&	3.9452\\
6.0$\times 10^{16}$&	-&	8.2139&	12.2221&	1.3791&	-&	1.3791&	-&	-&	4.0679\\
1.0$\times 10^{17}$&	-&	8.1649&	12.1958&	1.3822&	-&	1.3822&	-&	-&	4.0771\\

\hline
\multicolumn{10}{c}{$\phantom{\Big(}\xi_0=0.1$}\\[2pt]
\hline \\[-15pt]
1.0$\times 10^{14}$&	0.4306&	12.2483&	12.3651&	0.1567&	0.1460&	0.1562&	-0.1406&	-0.0049&	0.4623\\
2.0$\times 10^{14}$&	0.4844&	13.2837&	13.6042&	0.3999&	0.1642&	0.3985&	-0.1576&	-0.0052&	1.1795\\
3.0$\times 10^{14}$&	0.5190&	13.6642&	14.2137&	0.6526&	0.1760&	0.6501&	-0.1680&	-0.0054&	1.9250\\
4.0$\times 10^{14}$&	0.5450&	13.7564&	14.5375&	0.8871&	0.1848&	0.8831&	-0.1755&	-0.0053&	2.6165\\
6.0$\times 10^{14}$&	0.5839&	13.5501&	14.7617&	1.2666&	0.1980&	1.2596&	-0.1858&	-0.0051&	3.7361\\
8.0$\times 10^{14}$&	0.6132&	13.1368&	14.7123&	1.5285&	0.2079&	1.5179&	-0.1925&	-0.0048&	4.5086\\
1.0$\times 10^{15}$&	0.6369&	12.6712&	14.5431&	1.6984&	0.2159&	1.6839&	-0.1970&	-0.0044&	5.0096\\
1.5$\times 10^{15}$&	0.6823&	11.5849&	13.9683&	1.8812&	0.2313&	1.8560&	-0.2026&	-0.0036&	5.5488\\
1.8$\times 10^{15}$&	0.7038&	11.0426&	13.6245&	1.9035&	0.2386&	1.8715&	-0.2035&	-0.0031&	5.6145\\
2.0$\times 10^{15}$&	0.7165&	10.7269&	13.4115&	1.9005&	0.2429&	1.8638&	-0.2033&	-0.0029&	5.6058\\
3.0$\times 10^{15}$&	0.7677&	9.5830&	12.5737&	1.8047&	0.2603&	1.7441&	-0.1976&	-0.0020&	5.3233\\
6.0$\times 10^{15}$&	0.8637&	8.3183&	11.5589&	1.5127&	0.2928&	1.3893&	-0.1684&	-0.0009&	4.4620\\
1.0$\times 10^{16}$&	0.9420&	8.1305&	11.3823&	1.3321&	0.3194&	1.1630&	-0.1497&	-0.0005&	3.9291\\
1.5$\times 10^{16}$&	1.0092&	8.2145&	11.3384&	1.2135&	0.3422&	1.0299&	-0.1581&	-0.0004&	3.5794\\
3.0$\times 10^{16}$&	1.1354&	7.6987&	10.2505&	0.9529&	0.3849&	0.8780&	-0.3094&	-0.0005&	2.8108\\
6.0$\times 10^{16}$&	1.2774&	5.6583&	7.4028&	0.7338&	0.4331&	1.7164&	-1.4149&	-0.0009&	2.1643\\
1.0$\times 10^{17}$&	1.3933&	4.4538&	5.6845&	0.7137&	0.4724&	5.9667&	-5.7242&	-0.0012&	2.1053\\

\hline
\multicolumn{10}{c}{$\phantom{\Big(}\xi_0=0.2$}\\[2pt]
\hline\\[-15pt]
1.0$\times 10^{14}$&	0.8612&	12.1022&	12.1922&	0.1492&	0.2920&	0.1472&	-0.2703&	-0.0197&	0.4400\\
2.0$\times 10^{14}$&	0.9689&	13.1271&	13.4107&	0.3805&	0.3285&	0.3748&	-0.3016&	-0.0213&	1.1222\\
3.0$\times 10^{14}$&	1.0380&	13.5051&	14.0074&	0.6206&	0.3519&	0.6104&	-0.3200&	-0.0217&	1.8306\\
4.0$\times 10^{14}$&	1.0900&	13.5984&	14.3220&	0.8431&	0.3695&	0.8275&	-0.3323&	-0.0217&	2.4867\\
6.0$\times 10^{14}$&	1.1678&	13.3986&	14.5326&	1.2021&	0.3959&	1.1744&	-0.3474&	-0.0208&	3.5457\\
8.0$\times 10^{14}$&	1.2263&	12.9932&	14.4724&	1.4481&	0.4158&	1.4068&	-0.3549&	-0.0195&	4.2714\\
1.0$\times 10^{15}$&	1.2738&	12.5352&	14.2933&	1.6057&	0.4318&	1.5498&	-0.3578&	-0.0182&	4.7363\\
1.5$\times 10^{15}$&	1.3646&	11.4616&	13.6905&	1.7671&	0.4627&	1.6731&	-0.3537&	-0.0150&	5.2122\\
1.8$\times 10^{15}$&	1.4076&	10.9216&	13.3251&	1.7796&	0.4772&	1.6626&	-0.3468&	-0.0134&	5.2492\\
2.0$\times 10^{15}$&	1.4330&	10.6050&	13.0949&	1.7707&	0.4858&	1.6386&	-0.3413&	-0.0125&	5.2228\\
3.0$\times 10^{15}$&	1.5353&	9.4222&	12.1273&	1.6464&	0.5205&	1.4455&	-0.3104&	-0.0093&	4.8561\\
6.0$\times 10^{15}$&	1.7273&	7.6148&	10.1804&	1.2588&	0.5856&	0.9359&	-0.2564&	-0.0063&	3.7130\\
1.0$\times 10^{16}$&	1.8840&	6.0863&	8.1143&	0.9990&	0.6387&	0.6391&	-0.2722&	-0.0067&	2.9466\\
1.5$\times 10^{16}$&	2.0185&	4.9563&	6.4447&	0.9069&	0.6843&	0.6076&	-0.3768&	-0.0082&	2.6751\\
3.0$\times 10^{16}$&	2.2709&	4.0842&	4.7288&	0.9409&	0.7699&	1.6547&	-1.4735&	-0.0101&	2.7754\\
6.0$\times 10^{16}$&	2.5549&	4.0725&	3.9632&	1.0863&	0.8662&	10.3705&	-10.1402&	-0.0102&	3.2041\\
1.0$\times 10^{17}$&	2.7867&	4.3200&	3.6586&	1.2319&	0.9448&	44.7764&	-44.4798&	-0.0095&	3.6336\\

\hline
\end{tabular}
\label{tab1}
\end{table}

\begin{table}
 \caption{Characteristics of a set of configurations for EOS2.
 The notations are the same as in Table \ref{tab1}.}
\vspace{.3cm}
\begin{tabular}{p{1.5cm}p{1.5cm}p{1.5cm}p{1.5cm}p{1.5cm}p{1.5cm}p{1.5cm}p{1.5cm}p{1.5cm}p{1.5cm}}
\hline \\[-5pt]
$\rho_{b c}, \text{g cm}^{-3}$ &  $ \hphantom{xx}R_{\text{th}},$ km & $ \hphantom{xx}R,$ km &$ R_{\text{prop}},$ km &
 $\hphantom{xx} M/M_\odot$ &  $\hphantom{x} M_{\text{th}}/M_\odot$&  $\hphantom{x} M_{\text{fl}}/M_\odot$
&  $M_{\text{sfint}}/M_\odot$&  $\hphantom{x} M_{\text{sfext}}/M_\odot$& $ \hphantom{xx} r_g,$ km \\[2pt]
\hline \\[-7pt]
\multicolumn{10}{c}{Without a wormhole} \\[2pt]
\end{tabular}
\begin{tabular}{p{1.5cm}.........}
\hline \\[-15pt]
1.0$\times 10^{13}$&	-&	35.1484&	35.3940&	0.2777&	-&	0.2777&	-&	-&	0.8192\\
3.0$\times 10^{13}$&	-&	34.0595&	34.7518&	0.7573&	-&	0.7573&	-&	-&	2.2338\\
5.0$\times 10^{13}$&	-&	33.0649&	34.1519&	1.1532&	-&	1.1532&	-&	-&	3.4014\\
7.0$\times 10^{13}$&	-&	32.1529&	33.5906&	1.4818&	-&	1.4818&	-&	-&	4.3708\\
1.0$\times 10^{14}$&	-&	30.9169&	32.8113&	1.8760&	-&	1.8760&	-&	-&	5.5335\\
2.0$\times 10^{14}$&	-&	27.6628&	30.6535&	2.6482&	-&	2.6482&	-&	-&	7.8112\\
3.0$\times 10^{14}$&	-&	25.3072&	28.9877&	2.9832&	-&	2.9832&	-&	-&	8.7993\\
4.0$\times 10^{14}$&	-&	23.5170&	27.6575&	3.1213&	-&	3.1213&	-&	-&	9.2068\\
6.0$\times 10^{14}$&	-&	20.9771&	25.6722&	3.1564&	-&	3.1564&	-&	-&	9.3101\\
8.0$\times 10^{14}$&	-&	19.2619&	24.2625&	3.0779&	-&	3.0779&	-&	-&	9.0787\\
1.0$\times 10^{15}$&	-&	18.0313&	23.2159&	2.9715&	-&	2.9715&	-&	-&	8.7647\\
1.5$\times 10^{15}$&	-&	16.1136&	21.5291&	2.7156&	-&	2.7156&	-&	-&	8.0099\\
2.0$\times 10^{15}$&	-&	15.0592&	20.5768&	2.5160&	-&	2.5160&	-&	-&	7.4212\\
3.0$\times 10^{15}$&	-&	14.1095&	19.7367&	2.2541&	-&	2.2541&	-&	-&	6.6487\\
6.0$\times 10^{15}$&	-&	14.0299&	19.8738&	1.9703&	-&	1.9703&	-&	-&	5.8117\\
1.0$\times 10^{16}$&	-&	14.8886&	20.9756&	1.9447&	-&	1.9447&	-&	-&	5.7360\\
1.5$\times 10^{16}$&	-&	15.5199&	21.8035&	2.0028&	-&	2.0028&	-&	-&	5.9076\\
3.0$\times 10^{16}$&	-&	15.8192&	22.3168&	2.1019&	-&	2.1019&	-&	-&	6.1998\\

\hline
\multicolumn{10}{c}{$\phantom{\Big(}\xi_0=0.1$}\\[2pt]
\hline \\[-15pt]
1.0$\times 10^{13}$&	1.1374&	35.0540&	35.2774&	0.2717&	0.3856&	0.2708&	-0.3727&	-0.0121&	0.8014\\
3.0$\times 10^{13}$&	1.1374&	33.9788&	34.6415&	0.7409&	0.3856&	0.7383&	-0.3714&	-0.0116&	2.1855\\
5.0$\times 10^{13}$&	1.1374&	32.9983&	34.0493&	1.1283&	0.3856&	1.1240&	-0.3701&	-0.0112&	3.3282\\
7.0$\times 10^{13}$&	1.1374&	32.0990&	33.4947&	1.4500&	0.3856&	1.4441&	-0.3688&	-0.0108&	4.2771\\
1.0$\times 10^{14}$&	1.1374&	30.8813&	32.7260&	1.8360&	0.3856&	1.8276&	-0.3669&	-0.0103&	5.4154\\
2.0$\times 10^{14}$&	1.1374&	27.6798&	30.6016&	2.5927&	0.3856&	2.5764&	-0.3606&	-0.0087&	7.6474\\
3.0$\times 10^{14}$&	1.1374&	25.3698&	28.9684&	2.9217&	0.3856&	2.8980&	-0.3544&	-0.0075&	8.6180\\
4.0$\times 10^{14}$&	1.1374&	23.6247&	27.6755&	3.0584&	0.3856&	3.0276&	-0.3482&	-0.0066&	9.0210\\
6.0$\times 10^{14}$&	1.1374&	21.1631&	25.7582&	3.0959&	0.3856&	3.0518&	-0.3363&	-0.0052&	9.1318\\
8.0$\times 10^{14}$&	1.1374&	19.5213&	24.4165&	3.0229&	0.3856&	2.9665&	-0.3249&	-0.0043&	8.9165\\
1.0$\times 10^{15}$&	1.1374&	18.3622&	23.4391&	2.9230&	0.3856&	2.8551&	-0.3141&	-0.0036&	8.6218\\
1.5$\times 10^{15}$&	1.1374&	16.6197&	21.9273&	2.6850&	0.3856&	2.5915&	-0.2896&	-0.0025&	7.9197\\
2.0$\times 10^{15}$&	1.1374&	15.7405&	21.1547&	2.5036&	0.3856&	2.3884&	-0.2686&	-0.0018&	7.3847\\
3.0$\times 10^{15}$&	1.1374&	15.1024&	20.6263&	2.2768&	0.3856&	2.1276&	-0.2353&	-0.0011&	6.7156\\
6.0$\times 10^{15}$&	1.1374&	15.7007&	21.3879&	2.0735&	0.3856&	1.8706&	-0.1823&	-0.0005&	6.1160\\
1.0$\times 10^{16}$&	1.1374&	17.0156&	22.7755&	2.0881&	0.3856&	1.8611&	-0.1583&	-0.0003&	6.1592\\
1.5$\times 10^{16}$&	1.1374&	18.1359&	23.8678&	2.1448&	0.3856&	1.9081&	-0.1486&	-0.0002&	6.3264\\
3.0$\times 10^{16}$&	1.1374&	19.8761&	25.3427&	2.1975&	0.3856&	1.9480&	-0.1359&	-0.0002&	6.4818\\

\hline
\multicolumn{10}{c}{$\phantom{\Big(}\xi_0=0.15$}\\[2pt]
\hline\\[-15pt]
1.0$\times 10^{13}$&	1.7061&	34.9308&	35.1277&	0.2659&	0.5784&	0.2640&	-0.5492&	-0.0273&	0.7844\\
3.0$\times 10^{13}$&	1.7061&	33.8677&	34.4974&	0.7252&	0.5784&	0.7193&	-0.5463&	-0.0263&	2.1391\\
5.0$\times 10^{13}$&	1.7061&	32.8960&	33.9080&	1.1043&	0.5784&	1.0946&	-0.5434&	-0.0253&	3.2572\\
7.0$\times 10^{13}$&	1.7061&	32.0063&	33.3577&	1.4190&	0.5784&	1.4056&	-0.5406&	-0.0244&	4.1856\\
1.0$\times 10^{14}$&	1.7061&	30.8018&	32.5950&	1.7965&	0.5784&	1.7776&	-0.5363&	-0.0232&	5.2990\\
2.0$\times 10^{14}$&	1.7061&	27.6376&	30.4894&	2.5357&	0.5784&	2.4993&	-0.5223&	-0.0197&	7.4792\\
3.0$\times 10^{14}$&	1.7061&	25.3584&	28.8739&	2.8559&	0.5784&	2.8033&	-0.5087&	-0.0171&	8.4239\\
4.0$\times 10^{14}$&	1.7061&	23.6387&	27.5959&	2.9876&	0.5784&	2.9197&	-0.4956&	-0.0150&	8.8123\\
6.0$\times 10^{14}$&	1.7061&	21.2228&	25.7085&	3.0202&	0.5784&	2.9243&	-0.4706&	-0.0119&	8.9084\\
8.0$\times 10^{14}$&	1.7061&	19.6208&	24.3938&	2.9446&	0.5784&	2.8236&	-0.4477&	-0.0098&	8.6853\\
1.0$\times 10^{15}$&	1.7061&	18.4970&	23.4397&	2.8426&	0.5784&	2.6990&	-0.4266&	-0.0083&	8.3846\\
1.5$\times 10^{15}$&	1.7061&	16.8313&	21.9741&	2.5989&	0.5784&	2.4079&	-0.3816&	-0.0058&	7.6657\\
2.0$\times 10^{15}$&	1.7061&	16.0135&	21.2281&	2.4094&	0.5784&	2.1817&	-0.3463&	-0.0044&	7.1067\\
3.0$\times 10^{15}$&	1.7061&	15.4550&	20.6941&	2.1579&	0.5784&	1.8796&	-0.2972&	-0.0029&	6.3648\\
6.0$\times 10^{15}$&	1.7061&	16.0124&	21.0850&	1.8223&	0.5784&	1.4805&	-0.2350&	-0.0016&	5.3750\\
1.0$\times 10^{16}$&	1.7061&	16.8904&	21.5848&	1.5973&	0.5784&	1.2302&	-0.2101&	-0.0012&	4.7113\\
1.5$\times 10^{16}$&	1.7061&	16.9757&	21.1650&	1.3620&	0.5784&	0.9774&	-0.1927&	-0.0012&	4.0173\\
3.0$\times 10^{16}$&	1.7061&	13.5704&	16.5958&	0.8878&	0.5784&	0.4483&	-0.1375&	-0.0014&	2.6186\\

\hline
\end{tabular}
\label{tab2}
\end{table}

\subsection{Pressure}
\label{sect_press}

In the presence of a scalar field,
the effective pressure becomes anisotropic
(see, e.g.,~\cite{Gleiser:1988rq} and references therein).
Indeed, as follows from Eq.~\eqref{emt_wh_star_poten},
the  radial  pressure  $p_r=-T_1^1$ is distinct from
the tangential pressure  $p_t=-T_2^2=-T_3^3$ of the system.
This allows to introduce the fractional anisotropy $fa$,
\begin{equation}
\label{fa}
fa\equiv (p_r-p_t)/p_r.
\end{equation}

For the star-plus-wormhole system, the fractional anisotropy is given by
$$
fa=\frac{\left[1-2\sigma(n+1)v/\xi\right]\phi^{\prime 2}}
{-\sigma  \theta^{n+1}+1/2 \left[1-2\sigma(n+1)v/\xi\right]\phi^{\prime 2}}\,.
$$
Thus, at the core $\xi_0$ of the configurations
the fractional anisotropy is
$fa(\xi_0)=2\left[2\sigma^2(n+1)\xi_0^2+1\right]$.
Beyond the surface of the fluid, i.e., when $\xi\geq\xi_b$,
the fractional anisotropy is always equal to two.
The above expression also implies
that the numerator never vanishes for $\xi>\xi_0$.
The denominator, however, might tend to zero at certain values
of the radius. This would give rise to a tremendous growth
of the fractional anisotropy in the vicinity of these radii.

A related interesting consequence is the fact
that the  pressure of the neutron fluid and the total internal pressure
of the system (which includes the pressure contributions
from the fluid and the scalar field) differ.
Indeed, for the neutron fluid the pressure $p$ is given by
Eq.~\eqref{pressure_fluid_theta}.
In turn, from  Eqs.~\eqref{sf_poten}, \eqref{pressure_fluid_theta},
\eqref{nu_app}, and \eqref{dimless_xi_v} in \eqref{T11_stat},
one finds for the radial internal pressure of the system
\begin{equation}
\label{pres_int}
p_{\text{int}}\equiv  -\left[T_1^1\right]_{\text{int}}=\rho_{b c} c^2
\left\{\sigma \theta^{n+1}
-\frac{1}{2}\frac{\bar{D}^2}{\xi^4}e^{-\nu_c}
\left[\frac{1+\sigma (n+1)\theta}{1+\sigma (n+1)}\right]^{2}
\right\}.
\end{equation}
Taking into account Eqs.~\eqref{bound_v} and \eqref{const_D},
the ratio of the total radial internal pressure \eqref{pres_int}
and the fluid pressure \eqref{pressure_fluid_theta}
at the core of the configuration is given by
\begin{equation}
\label{pres_ratio}
\frac{p_{\text{int}}}{p}=-\frac{1}{2\sigma^2(n+1)\xi_0^2}.
\end{equation}
In principle, for large values of $\xi_0$ and $\sigma$
the absolute value of this ratio can be small.
However, for the configurations of most interest to us,
i.e., for the potentially stable configurations, $\sigma$ is less than one.
Then, Eq.~(52) shows that the absolute value of this ratio is typically
considerably greater than one
when considering also configurations with $\xi_0$ less than one.
Obviously, the difference in these pressures
arises because of the presence of the ghost scalar field.

\subsection{Numerical results}
\label{num_res}

\begin{figure}[t]
\begin{minipage}[t]{.49\linewidth}
  \begin{center}
  \includegraphics[width=7cm]{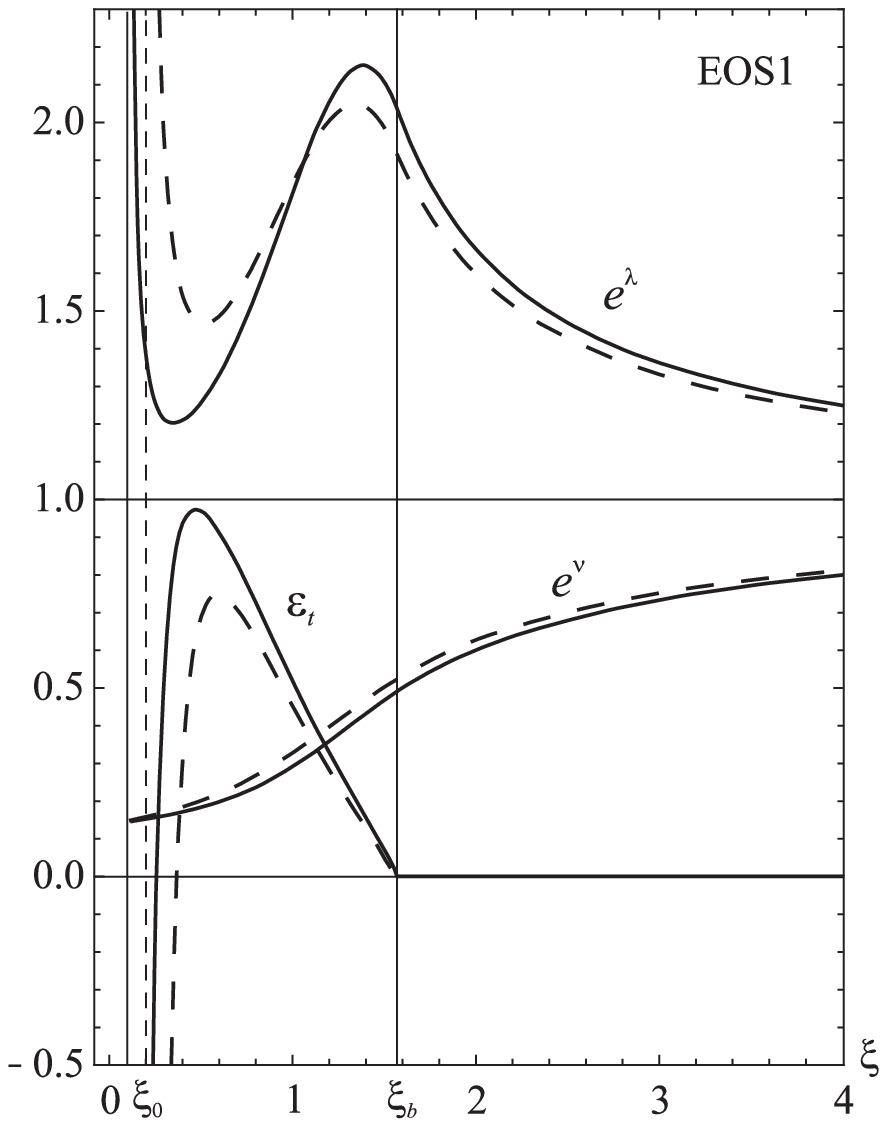}
  \end{center}
\end{minipage}\hfill
\begin{minipage}[t]{.49\linewidth}
  \begin{center}
  \includegraphics[width=7cm]{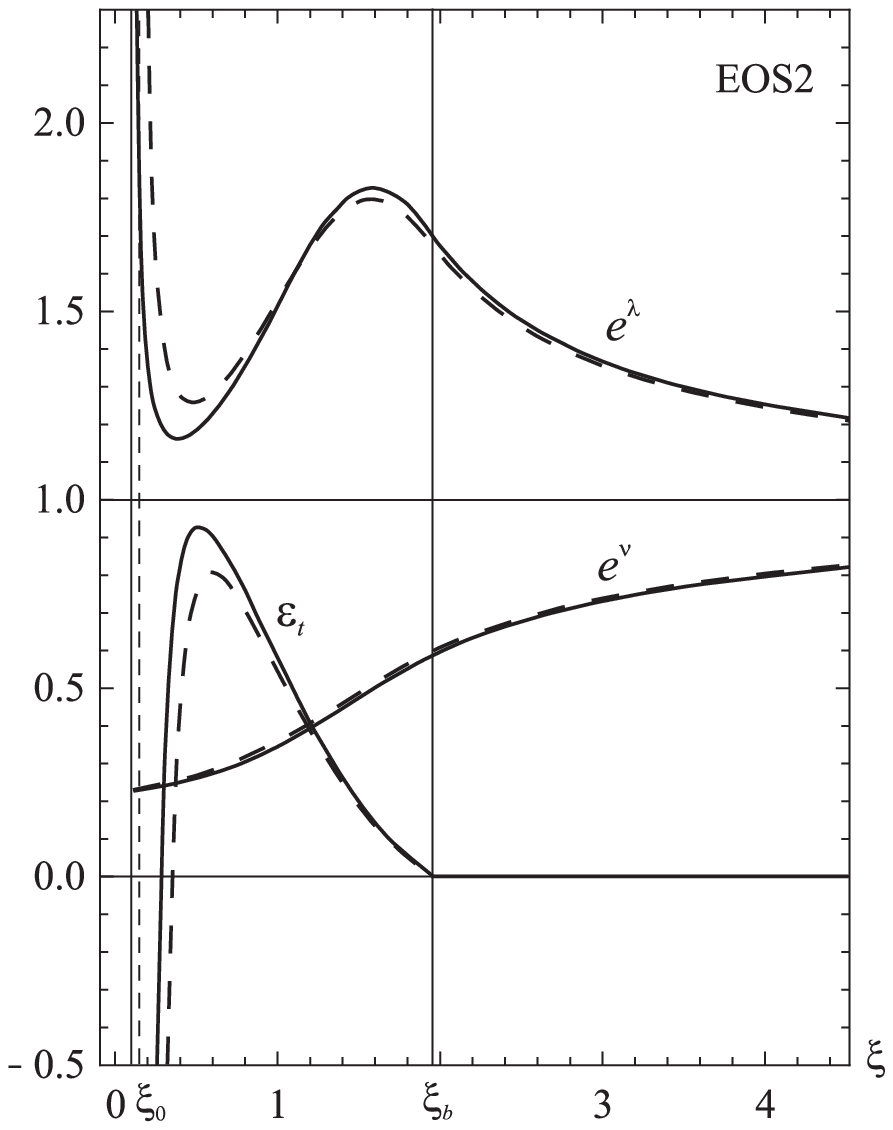}
  \end{center}
\end{minipage}\hfill
  \caption{The metric functions $g_{tt}=e^{\nu}$, $g_{rr}=-e^{\lambda}$,
and the total energy density $\varepsilon_t$
from \eqref{dens_int} and \eqref{dens_ext}
(in units of $\rho_{b c} c^2$)
for two equations of state EOS1 (left panel)
and EOS2 (right panel).
The graphs are shown for configurations with the maximum mass
$M$ according to Tables \ref{tab1} and \ref{tab2}.
The solid and the dashed curves correspond
to $\xi_0=0.1$ and $\xi_0=0.2$ for EOS1,
and $\xi_0=0.1$ and $\xi_0=0.15$ for EOS2, respectively.
The thin vertical line marked by $\xi_b$
corresponds to the boundary of the fluid
where $\theta=0$.
The thin vertical lines marked by $\xi_0$ correspond to
the core of the respective configurations.
The values of the total energy density at the core are
for EOS1 $\varepsilon_t(\xi_0=0.1) \approx -59.34$
and $\varepsilon_t(\xi_0=0.2) \approx -14.17$;
for EOS2 $\varepsilon_t(\xi_0=0.1) \approx -82.00$
and $\varepsilon_t(\xi_0=0.15) \approx -35.89$.
To provide asymptotic flatness of the solutions,
the value of the constant $\nu_c$ is chosen
for EOS1 at $\xi_0=0.1$ $\nu_c \approx -1.919$,
at $\xi_0=0.2$ $\nu_c \approx -1.865$;
for EOS2 at $\xi_0=0.1$ $\nu_c \approx -1.472$,
at $\xi_0=0.15$ $\nu_c \approx -1.455$.
}
\label{energ_metr_fig}
\end{figure}

\begin{figure}[h!]
\centering
  \includegraphics[height=10.7cm]{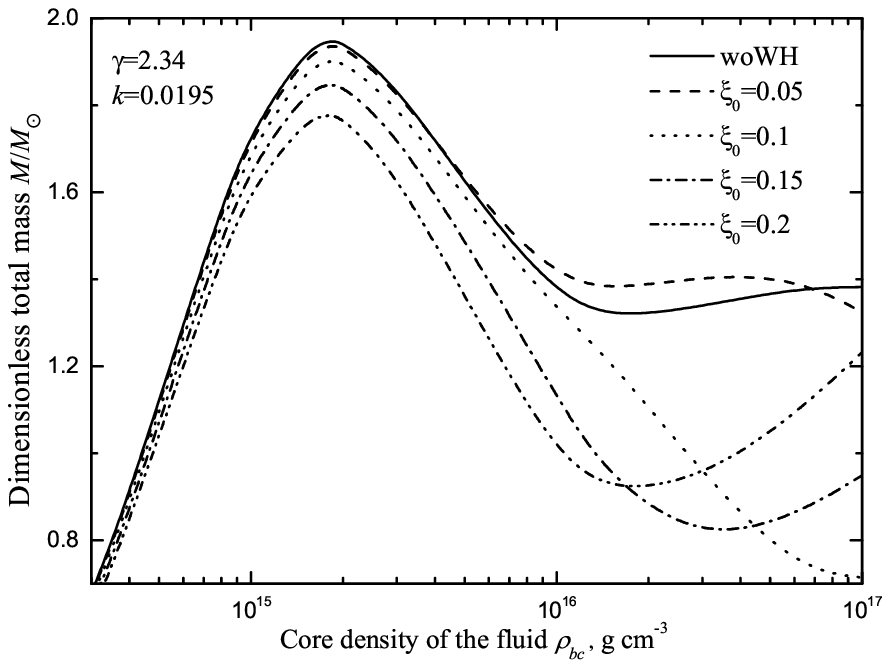}
  \includegraphics[height=10.7cm]{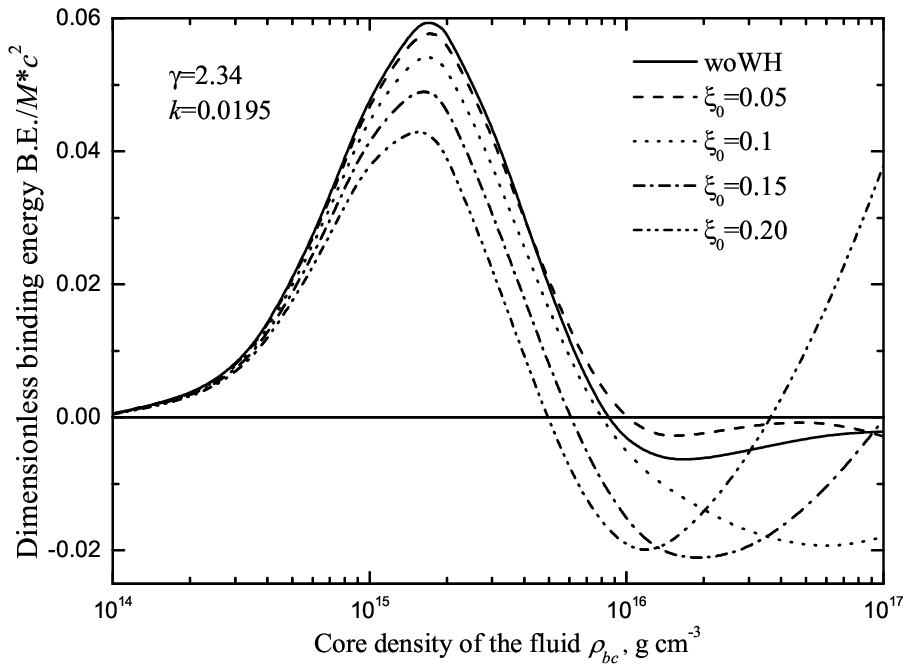}
\vspace{-1.cm}
\caption{The case of EOS1:\\
(a) Total mass of the configurations (in solar mass units), $M/M_\odot$,
versus the core density $\rho_{b c}$ for various values
of the dimensionless throat radius $\xi_0$.
Stable configurations should reside to the left of the first mass peak.
\\
(b) The dimensionless binding energy given by Eq.~\eqref{bind_enrg}
versus the core density $\rho_{b c}$ for
various values of the dimensionless throat radius $\xi_0$.
}
\label{fig_mass_total_234}
\end{figure}

\begin{figure}[h!]
\centering
  \includegraphics[height=10.7cm]{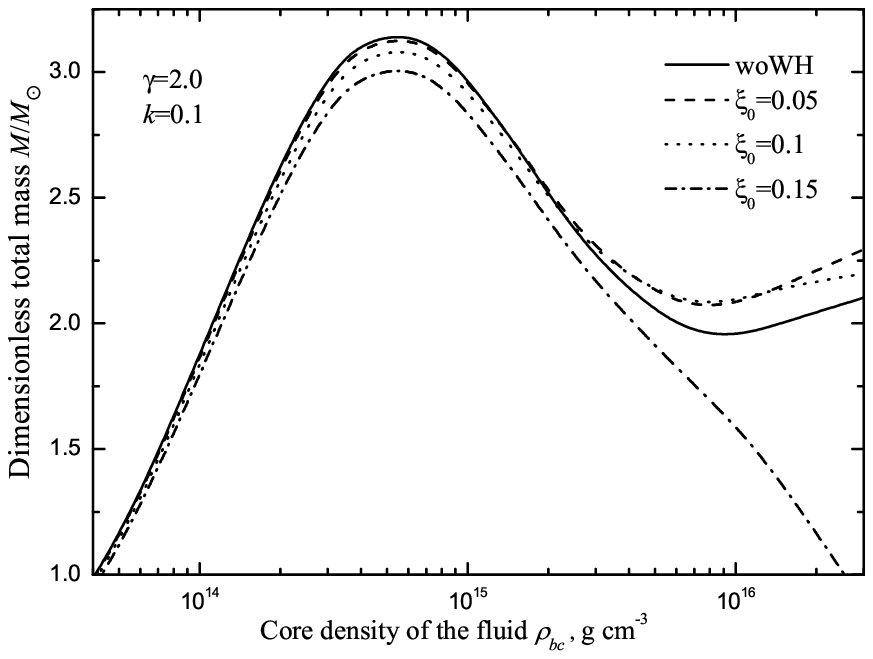}
  \includegraphics[height=10.7cm]{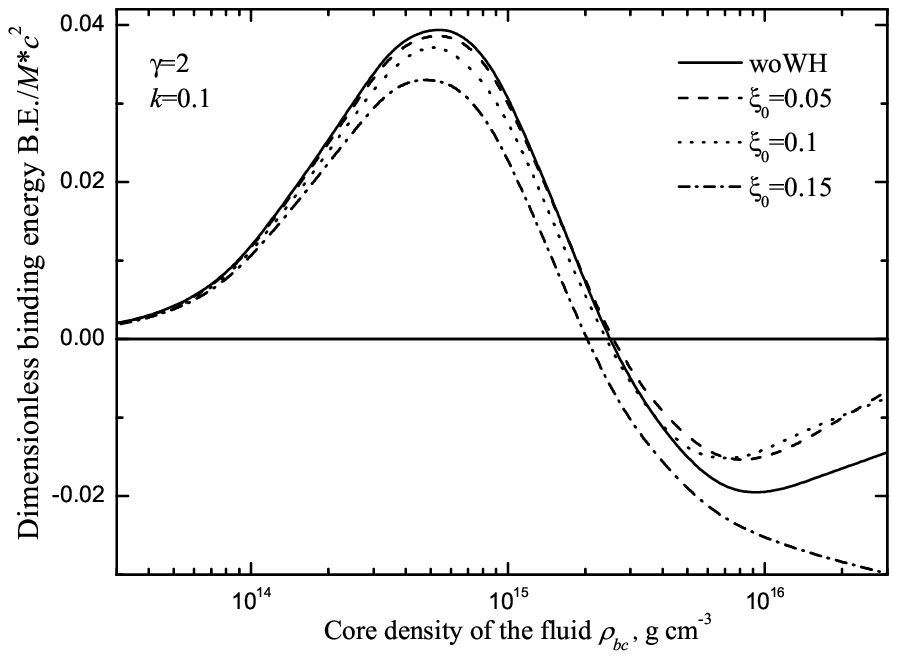}
\vspace{-1.cm}
\caption{The case of EOS2:\\
(a) Total mass of the configurations (in solar mass units), $M/M_\odot$,
versus the core density $\rho_{b c}$ for various values
of the dimensionless throat radius $\xi_0$.
Stable configurations should reside to the left of the first mass peak.
\\
(b) The dimensionless binding energy given by Eq.~\eqref{bind_enrg}
versus the core density $\rho_{b c}$ for
various values of the dimensionless throat radius $\xi_0$.
}
\label{fig_mass_total_200}
\end{figure}

Examples of numerical solutions of the sets of equations
subject to the appropriate sets of boundary conditions
are presented in Fig.~\ref{energ_metr_fig}.
In particular, we exhibit
the metric functions $g_{tt}=e^{\nu}$ and $g_{rr}=-e^\lambda$
and the distribution of the total energy density $\varepsilon_t$
of the system consisting of the scalar-field energy density and
the fluid energy density.
For the graphs, we have determined
$e^{\nu}$ from Eq.~\eqref{nu_app}, and
$e^\lambda$ follows from Eq.~\eqref{u_app}
together with \eqref{dimless_xi_v}
\begin{equation}
\label{lambda_poly}
e^\lambda=\left[1-2\sigma(n+1)\frac{v}{\xi}\right]^{-1}.
\end{equation}
(Note, that this expression diverges at $\xi=\xi_0$,
as discussed in connection with the boundary conditions.)
The total energy density $\varepsilon_t$ is determined
according to Eq.~(\ref{total_edens}),
and consists of the internal energy density
\eqref{dens_int} and the external energy density \eqref{dens_ext}.
As seen from Fig.~\ref{energ_metr_fig},
at the boundary $\xi=\xi_b$ of the fluid
the total energy density is almost vanishing,
since the scalar-field ``tail'' beyond $\xi_b$
gives only a very small contribution to the total energy density.

We exhibit a number of characteristic properties
of the star-plus-wormhole systems in Tables \ref{tab1} and  \ref{tab2}
for the two equations of state EOS1 and EOS2, respectively.
Choosing $\xi_0=0$, we start in each case
with a sequence of ordinary neutron stars for later reference.
We then consider sequences of star-plus-wormhole systems
for several fixed finite values of $\xi_0$.
Each sequence of configurations is obtained by increasing the
core density of the fluid $\rho_{b c}$
in a physically relevant range of values.

The tables then exhibit for each configuration the
areal radius of the throat $R_{\rm th}$, the
areal radius of the fluid $R$,
the proper radius of the fluid $R_{\rm prop}$
and the gravitational radius of the system $r_g$,
all in units of kilometers.
In addition, the total mass $M$,
and the mass contributions
$M_{\text{th}}$, $M_{\text{fl}}$, $M_{\text{sfint}}$ and $M_{\text{sfext}}$
are presented in the Tables, employing solar mass units.
(In order not to encumber the paper,
only a few values of $\xi_0$ have been selected in the Tables.)

The dependence of the total mass (in solar mass units)
on the core density $\rho_{b c}$ (in grams per cubic centimeter)
is shown in
Figs.~\ref{fig_mass_total_234}(a) and \ref{fig_mass_total_200}(a)
for the two equations of state, respectively,
and several values of the core radius $\xi_0$.
The corresponding values of the binding energy
Eq.~\eqref{bind_enrg} are shown in
Figs.~\ref{fig_mass_total_234}(b) and \ref{fig_mass_total_200}(b).
In these figures, the curves labelled by ``wo WH''
correspond to the configurations without a wormhole
(and correspondingly without a scalar field),
representing ordinary neutron stars
modeled by the equation of state \eqref{eqs_NS_WH}.

We note that as a function of the core density $\rho_{b c}$,
the total mass $M$ of the configurations rises
monotonically to a maximum and then decreases again.
This is typical for this type of configurations.
In Tooper's paper \cite{Tooper2}, for instance,
ordinary neutron stars were investigated in detail
for various values of the polytropic index $n$.
In particular, it was shown that
the first peak in the mass
corresponds to the point dividing stable and unstable
neutron-star configurations.
This first mass peak is reached
at a critical value of the core density,
$\rho^{(cr)}_{b c}$.

For the equations of state EOS1 and EOS2 used here,
the values of $n$ are $n\approx 0.75$ and $n= 1$, respectively.
Here, the typical behavior of the mass function
of the case without a wormhole persists
in the presence of the wormhole.
Thus, while the first mass peak persists,
the inclusion of a wormhole leads to a decrease
of the height of the mass peak with increasing wormhole throat.
This decrease is stronger for EOS1 than for EOS2.

A necessary, but not sufficient, condition
for the stability of the configurations
is the positivity of the binding energy.
Configurations with a negative binding energy are unstable
against dispersal of the matter to infinity.
In the case without a wormhole considered in \cite{Tooper2},
the binding energy always has a positive first peak for $n\lesssim 3$.
Moreover, it remains always positive to the left of the peak,
and becomes negative to the right of the peak
at some large value of $\rho_{b c}$.
Configurations to the left of the first peak are stable,
while configurations to the right are unstable,
because it is energetically favorable for them
to make a transition to a state with the same particle number $N$,
but with a smaller central density $\rho_{b c}$.

For the star-plus-wormhole systems, we expect an
analogous behaviour concerning their stability.
To the left of the peak, the solutions should be
energetically stable, while to the right of the peak
instability would set in.
This conclusion is supported by a stability analysis
via catastrophe theory, as often applied to boson stars
(see, e.g.,~\cite{Kusmartsev:2008py,Tamaki:2010zz,Kleihaus:2011sx}).
Here, one selects an appropriate set of
\textit{behavior variable(s)} and \textit{control parameter(s)}.
In our case, we could choose
the size of the wormhole throat $\xi_0$ and the mass $M$
as two control parameters,
and the core density $\rho_{b c}$ as a behavior variable.
According to catastrophe theory,
the stability with respect to local perturbations
then changes only at turning points,
where, in this case, ${\partial M}/{\partial \rho_{b c}} = 0$
for fixed $\xi_0$.
Thus, stability should change at the maximum of the mass.
However, a conclusive answer on the question of stability
should be obtained from a detailed stability mode analysis
of the star-plus-wormhole systems.

\section{Conclusions}
\label{conclus_WH_NS}

In this work, we have continued the investigations
of star-plus-wormhole configurations suggested in \cite{arXiv:1102.4454}.
These objects have a nontrivial wormhole-like topology
with a tunnel filled by ordinary matter.
For obtaining the nontrivial topology,
we have used one of the simplest forms of matter violating
the weak/null energy conditions -- a massless ghost scalar field.
The matter filling the wormhole has been taken
in the form of neutron matter approximated by an equation of state
of the form \eqref{eqs_NS_WH}.

Since the resulting star-plus-wormhole configurations
possess properties both of wormholes and of ordinary stars,
our goal has been to clarify
the influence of the presence of a ghost scalar field
on the physical characteristics of such well-studied configurations
as neutron stars.
Our main emphasis has been
(i) the study of the dependence of the mass
of the mixed star-plus-wormhole configurations
on the core density and on the size of the wormhole throat;
(ii) the study of the binding energy and the pressure
of the mixed star-plus-wormhole configurations.

Our results can be summarized as follows:

\vspace{-0.2cm}
\begin{enumerate}
\itemsep=-0.2pt
\item[(1)] 
For two polytropic equations of state denoted by EOS1 and EOS2
[see Eq.~\eqref{eqs_NS_WH} below],
sequences of static, regular solutions
describing neutron-star-plus-wormhole systems
have been constructed numerically,
by solving the coupled Einstein-matter equations
subject to appropriate boundary conditions.
Examples of such solutions are shown in Fig.~\ref{energ_metr_fig}.
\item[(2)] 
For the above equations of state,
the physical properties of the configurations are characterized
by two parameters~-- the core density of the fluid
and the dimensionless throat radius.
We have evaluated the total mass of the respective configurations,
and also the mass contributions of the individual components
the system consists of.
Examples of their values are given in Tables \ref{tab1} and \ref{tab2}.
Likewise, the tables exhibit the values of various physically
relevant radii of these configurations.
\item[(3)] 
As a function of the core density of the fluid,
the total mass of the star-plus-wormhole configurations rises
monotonically to a maximum and then decreases again,
as is typical also for ordinary neutron stars.
Since neutron stars located to the left of the maximum are stable,
we have conjectured that the same may hold for
the star-plus-wormhole configurations constructed here.
\item[(4)] 
In the region where stability may be possible,
the masses of configurations with a wormhole are always less
than the masses of ordinary neutron stars without a scalar field.
Correspondingly, the maximum values for the mass
of star-plus-wormhole configurations
are always less than for configurations without a wormhole present,
as seen in Figs.~\ref{fig_mass_total_234} and \ref{fig_mass_total_200}.
\end{enumerate}

The differences found between the characteristic properties
of neutron matter configurations with and without
a ghost scalar field, of course, are not the only ones
that are possible.
Further examples may include the following phenomena:

\begin{itemize}
\itemsep=-0.5pt
\item[(i)] The presence of an intrauniverse wormhole may lead
to the fact that a distant observer will see two stars
associated with the two mouths of the wormhole,
having the same or very similar characteristics.
\item[(ii)] Since the neutron matter can, in principle,
move freely through the tunnel,
this may lead to oscillations of matter
relative to the core of the configuration.
As a result, at various moments of time
there will be different quantities of matter
near the two mouths of the wormhole.
From the point of view of a distant observer,
this will look like periodical changes in the mass of the stars
and fluctuations of their luminosity.
\item[(iii)] The neutron-star-plus-wormhole configurations
considered here might possibly possess favorable conditions
to convert neutron matter into quark (or strange) matter,
conjectured to be present in quark stars
\cite{Itoh:1970uw,Witten:1984rs,Farhi:1984qu,Alcock:1986hz}.
The quark matter would not be able to decay into another
more energetically favorable state,
since it is assumed to possess the highest binding energy possible.
In this case, it might be possible that the final stage of evolution
of such configurations will be quark star-plus-wormhole configurations.
Such systems should have a number of distinctions as compared to
neutron-star-plus-wormhole configurations \cite{Alcock:1986hz}.
\item[(iv)] A further possible observationally interesting effect
of the presence of dark energy in mixed neutron-dark-energy configurations
consists in possible changes in the gravitational wave spectrum
as considered in Ref.~\cite{Yazadjiev:2011sd}.
The investigations performed there for configurations
with a trivial topology can be also done for systems
with a nontrivial topology as the ones studied here.
The consideration of these and similar issues
will be the object of further studies.
\end{itemize}

Let us finally come back to the question of stability of the
star-plus-wormhole systems.
It is well known, that an isolated wormhole with a massless ghost
scalar field is unstable
\cite{Shinkai:2002gv,Gonzalez:2008wd,Gonzalez:2008xk,Bronnikov:2011if}.
When filling such a wormhole with neutron matter,
a stable configuration may possibly result.
However, if a careful mode analysis will reveal that the
instability persists, one may turn to other, stable types of wormholes
when considering star-plus-wormhole systems.
Possible candidates for such wormholes are, for instance,
the recently constructed dilatonic Einstein-Gauss-Bonnet wormholes
\cite{Kanti:2011jz,Kanti:2011yv}.

\vspace{0.3cm}
\section*{Acknowledgements}
V.D. and V.F. are grateful to the Research Group Linkage
Programme of the Alexander von Humboldt Foundation for the
support of this research. They also would like to thank the
Carl von Ossietzky University of Oldenburg for hospitality
while this work was carried out.
This work is partially supported by a grant in fundamental research in natural
sciences by the Ministry of Education and Science of Kazakhstan.
B.K. and J.K. acknowledge support by  the~DFG.

\end{document}